\documentclass[graybox]{svmult}

\usepackage{type1cm}        
\usepackage{makeidx}         
\usepackage{graphicx}        
\usepackage{multicol}        
\usepackage[bottom]{footmisc}

\usepackage{newtxtext}       %
\usepackage{newtxmath}       

\makeindex             

\begin{document}

\title*{The D-CTC Condition in Quantum Field Theory}
\author{Rainer Verch}
\institute{Rainer Verch \at Institut f\"ur Theoretische Physik, Universit\"at Leipzig, D-04103 Leipzig, Germany. \email{rainer.verch{@}uni-leipzig.de}}
%
%
\maketitle


\abstract{A condition proposed by David Deutsch to describe analogues of processes in the presence of closed timelike curves (D-CTC condition) in bipartite quantum
systems is investigated within the framework of local relativistic quantum field theory. The main result is that in relativistic quantum field theory on spacetimes
where closed timelike curves are absent, the D-CTC condition can nevertheless be fulfilled to arbitrary precision, under very general, model-independent conditions. 
Therefore, the D-CTC condition should not be taken
as characteristic for quantum processes in the presence of closed timelike curves in the sense of general relativity. This report is a very condensed extract of the 
publication \cite{jtrv}.
A new result showing that the D-CTC condition can be approximately fulfilled by entangled states is also presented. 
}


\section{The D-CTC condition: Bipartite quantum systems}

The D-CTC condition has been introduced in 1991 in a seminal article by David Deutsch \cite{Deutsch}
as a condition for  processes involving an analog of closed timelike curves (CTCs) in
quantum circuits. Such a process in a quantum circuit is symbolically depicted in Fig.\ 1.
In more concrete terms, the simplest form of a quantum circuit is a
bipartite quantum system with subsystem Hilbert spaces $\mathcal{H}_A$ and $\mathcal{H}_B$
and $\mathcal{H} = \mathcal{H}_A \otimes \mathcal{H}_B$ as the total Hilbert space of the system,
together with a unitary operator $U : \mathcal{H} \to \mathcal{H}$ describing the dynamics (``one-time-step
time evolution'') of the 
system, coupling the two subsystems.
$-T$ symbolizes a ``time-step backward in time'', meaning that the partial state of full system 
on the system part $B$ after applying $U$ is the same as before applying $U$ on
system part $B$.

\begin{center}
 \includegraphics[width=8cm]{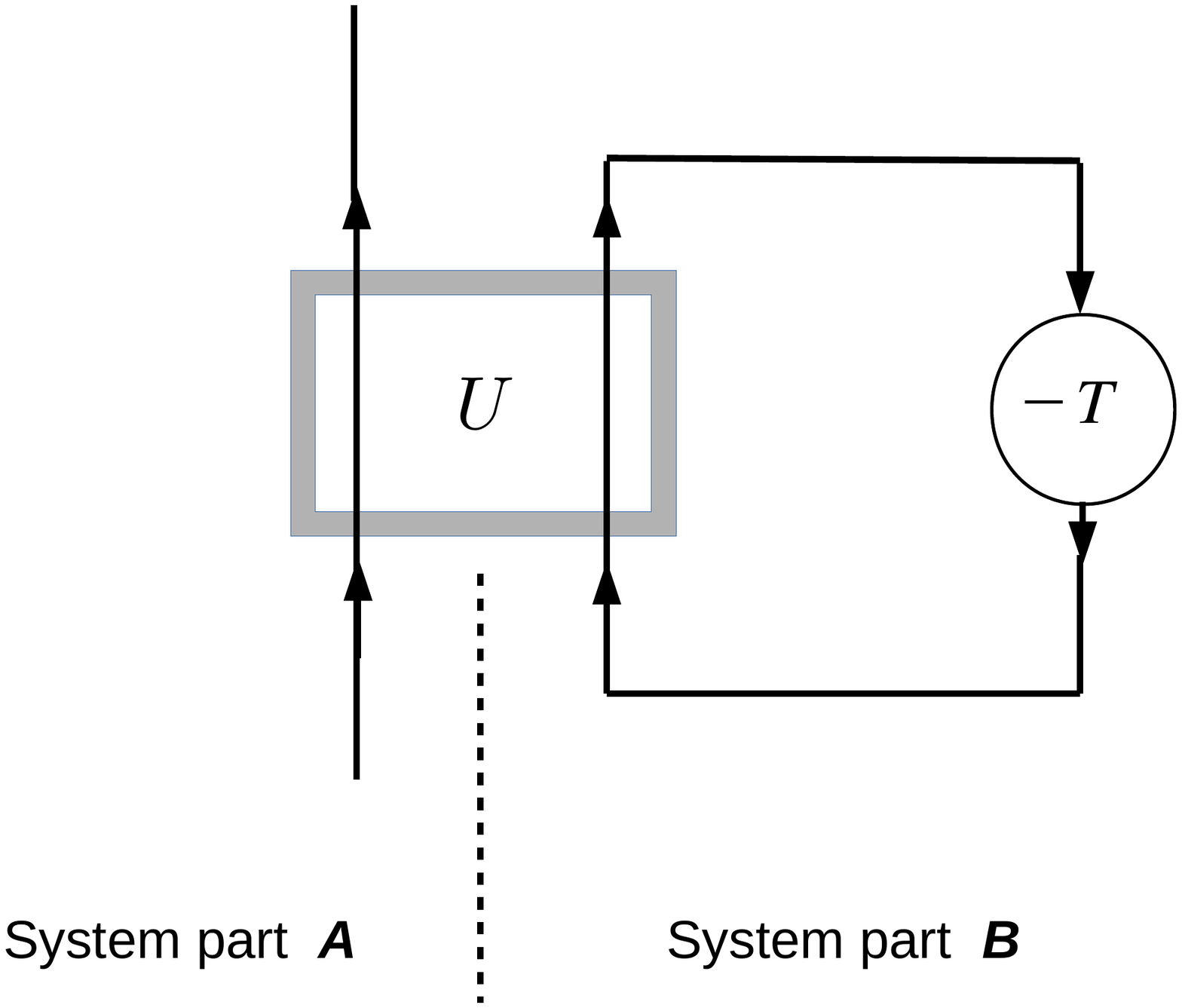}
\end{center}
{\small {\bf Figure 1.} A process in a quantum circuit is represented a unitary operator $U$
describing the dynamical coupling of two system parts (denoted by $A$ and $B$). A ``step backward in time''
is symbolized by $-T$; the $B$-part of the result of the process is again fed into the process
as initial state of the $B$-part.}
\\[6pt]
More formally, Deutsch's condition on quantum circuits with processes that provide ``analogues of closed time curves'',
that means, involving (dynamical) ``backward time steps'' --- referred to as {\it D-CTC condition} --- is verbally described as follows:
Given a unitary $U$ on $\mathcal{H}$ and a partial state (density matrix) $\varrho_A$ on system part $A$, 
a state (density matrix) $\varrho$ of the full system is said to fulfill the D-CTC condition if
the restriction of $\varrho$ to system part $A$ coincides with $\varrho_A$ and if $U\varrho U^\ast$ and $\varrho$ agree
when restricted to system part $B$. At the level of equations, this is expressed in the following way:
\begin{itemize}
 \item Given:  $U$ unitary on $\mathcal{H}$, \ \ $\varrho_A$ density matrix on $\mathcal{H}_A$
\end{itemize}
A density matrix $\varrho$ on $\mathcal{H}$ {\it fulfills the D-CTC condition} if
\begin{itemize}
 \item ${}$ \ \ ${\rm Tr}_B \varrho = \varrho_A$ \ \  $\Leftrightarrow {\rm Tr}(\varrho({\bf a} \otimes {\bf 1})) = {\rm Tr}_{\mathcal{H}_A}(\varrho_A {\bf a})$ \quad and
 \item ${}$ \ \ ${\rm Tr}_A U\varrho U^\ast = {\rm Tr}_A \varrho$ \ \ $\Leftrightarrow {\rm Tr}(\varrho({\bf 1} \otimes {\bf b})) =
 {\rm Tr}(U\varrho U^\ast({\bf 1} \otimes {\bf b}))$
\end{itemize}
${}$ \quad \ for all ${\bf a} \in {\sf B}(\mathcal{H}_A)$ and ${\bf b} \in {\sf B}(\mathcal{H}_B)$.
\\[6pt]
Deutsch has shown that there is always a solution to the problem of finding a density matrix state $\varrho$ fulfilling the 
D-CTC condition {\it if the Hilbert spaces $\mathcal{H}_A$ and $\mathcal{H}_B$ are finite-dimensional}:
\\[6pt]
Given $U$ and $\varrho_A$ = density matrix on $\mathcal{H}_A$, there is $\varrho_B$ = density matrix on $\mathcal{H}_B$ such that
$\varrho = \varrho_A \otimes \varrho_B$ fulfills the D-CTC condition. 
\\[6pt]
According to Deutsch, the proof is based on a fixed point argument: The map 
$$ \mathcal{S}: \varrho_B \mapsto {\rm Tr}_A( U(\varrho_A \otimes \varrho_B)U^\ast) $$
possesses fixed points.
\\[6pt]
The deeper reason why the result holds --- and why a fixed point argument can be used --- is that for quantum mechanical
systems with
finite dimensional Hilbert spaces, the
state space is convex and closed.

\section{The D-CTC Condition and Dynamics on CTC Spacetimes}
Spacetimes with CTCs are possible in General Relativity. Their physical realization as solutions to Einstein's equations
of gravity and matter is doubtful. The common understanding is that 
unusual matter properties (e.g.\ negative energy densities)
are required in order for spacetimes with CTCs to occur in general relativity, i.e.\ as solutions to Einstein's equations, and that
this is not feasible for macroscopic matter \cite{Hawking}.
\\[6pt]
Deutsch \cite{Deutsch} has proposed that quantum circuits with backward time-steps can be seen as analogues of dynamical (quantum) systems
on spacetimes with CTCs. The paradigm example to this end is the 2D (Deutsch-) Politzer spacetime which arises by cutting out two 
finite spacelike strips separated by a finite time segment and by identifying the ``inner rims'' and ``outer rims'' of the cuts as 
depicted in Fig.\ 2. For further discussion, see the refs.\ \cite{Politzer,CGS,FewsHiguWells}.
\begin{center}
 \includegraphics[width=7.0cm]{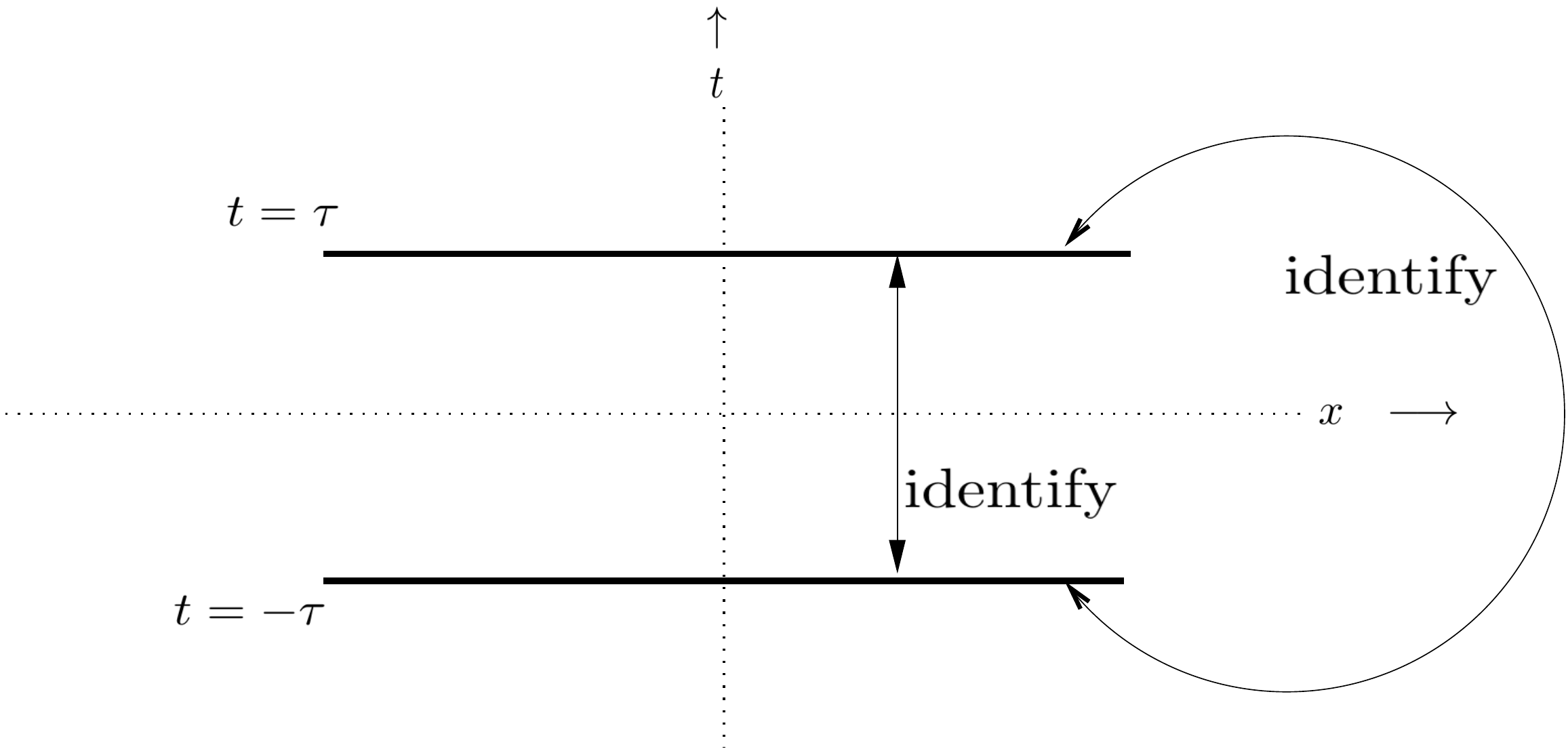}
\end{center}
{\small {\bf Figure 2.} The Politzer spacetime arises by cutting two spacelike line segments out of 1 + 1 dimensional 
Minkowski spacetime and by suitably identifying the cuts.}
\\[6pt]
One can consider (possibly multi-component)
wave functions $\psi(t,x)$ on Politzer spacetime fulfilling a hyperbolic wave equation which may be brought into the form
$$ \partial_t \psi(t,x) = D_x \psi(t,x) $$
where $D_x$ is a suitable (possibly matrix-formated) differential operator acting with respect to the $x$ coordinate.
(E.g.\ for a Klein-Gordon field $\phi(t,x)$, one would choose $\psi(t,x) = \left( {}^{\phi(t,x)}_{\partial_t \phi(t,x)} \right)$\,.)
The identification of the cuts in the Politzer spacetime imposes boundary conditions like
 $$ \lim_{\epsilon \to 0+} \,\psi(-\tau + \epsilon,x) - \psi(\tau - \epsilon,x) = 0$$
 along the cut line, and one may regard them as 
analogous to Deutsch's condition
 $${\rm Tr}_A(U \varrho U^\ast) = {\rm Tr}_A(\varrho)$$
 on viewing $U$ as analogous to the ``time-evolution'' 
 $$ \psi(-\tau,\,.\,) \to \psi(\tau,\,.\,) \,. $$
It is basically this analogy that serves as motivation to interpret the D-CTC condition for bipartite systems stated above
as representing processes in the presence of closed timelike curves. This proposal by Deutsch has stimulated several investigations
of the D-CTC condition including experimental tests; the references \cite{AhnMyersRalphMann,BrunWW,BubStairs,PienaarRalphMyers,RingbauerEtAl}
are only a few examples.
However, authors of more recent publications went beyond Deutsch's original interpretation of the D-CTC condition
which was mainly at the level of an analogy. E.g.\ in \cite{RingbauerEtAl} it is claimed that bipartite quantum systems
in which the D-CTC condition is fulfilled have experimentally been constructed, and that 
{\sl ``...quantum mechanics therefore allows for causality violation without paradoxes whilst remaining consistent with
relativity''}.
Taken literally, this amounts to a much stronger interpretation of the D-CTC condition, basically stating that it is 
characteristic of quantum processes in the presence of closed timelike curves in the sense of general relativity. From
our perspective, the question now arises if this stronger interpretation is justified, and how one could attain
an answer to that question --- bearing in mind that, apart from the analogy with fields on the Politzer spacetime, the 
D-CTC condition in its original formulation refers to general bipartite quantum systems without an a priori given relation
to spacetime structure. The connection between quantum physics and special or general relativity is provided by
relativistic quantum field theory, brought to the fore at its best in the local, operator algebraic approach \cite{Haag}. 
In fact, in \cite{jtrv} it is shown that the quantized massless scalar field can be constructed on the Politzer spacetime in
compliance with the principles of isotony of locality, upon making suitable adaptations due to the occurrence of closed timelike 
curves, like the F-locality concept due to Kay \cite{KayFlocality}. We refer to the references for full discussion, on which
we touch briefly in the following section.

\section{Relativistic Quantum Field Theory}

Relativistic quantum field theory, especially in the local, operator algebraic framework, addresses the localization of
observables and processes in the sense of special relativity, or in the sense of general relativity when suitably extending
the framework as quantum field theory in curved spacetimes supplied with local covariance \cite{Haag,RecAdvQFT}. Basic elements of 
the such extended framework, describing a quantum field theory on a $d+1$ dimensional Lorentzian spacetime $M$ with metric
$g$ and taken to be orientable and time-orientable, are: A quantum field theory on $M$ is given by 
a collection 
 $\mathcal{A}(O)$ of operator algebras indexed by subsets 
$O$ of $M$. Hermitean elements in $\mathcal{A}(O)$ represent observables which can be measured within $O$, and therefore,
the collection of operator algebras is subject to the following conditions,
\begin{itemize}
 \item Isotony: $O_1 \subset O_2$ $\Rightarrow$ $\mathcal{A}(O_1) \subset \mathcal{A}(O_2)$ 
 \item Locality: $O_1$ and $O_2$ causally separated $\Rightarrow$ $\mathcal{A}(O_1)$ and $\mathcal{A}(O_2)$ commute 
\end{itemize}
There are examples where the subsets $O$ of $M$ are a priori ``extended'' \cite{LongoMoriRehren,Rehren}. If the $\mathcal{A}(O)$ are ``large'' for
 arbitrarily ``small'' neighbourhoods of points in $M$, the QFT is called {\it arbitrarily localizable} (standard case if
 the $\mathcal{A}(O)$ derive from a Wightman quantum field).
\\[6pt]
If the spacetime $M$ with metric $g$ is {\it globally hyperbolic}, implying the absence of CTCs,
then this works fine; and there are many examples. In fact there is 
reason to assume that every physically realistic QFT on a globally hyperbolic spacetime complies with this 
basic structure. (See \cite{Fewster-Verch-Review} for discussion.)
\\[6pt]
If a spacetime contains CTCs, this is far from clear. 
\begin{itemize}
 \item In general, constraints stemming from CTCs could make the $\mathcal{A}(O)$ small 
 \item Isotony and Locality become highly non-trivial in presence of CTC constraints --- arbitrary localizability 
 won't hold in general 
\end{itemize}
In other words, the most basic conditions or principles generally assumed for observable quantities of quantum field theories
on globally hyperbolic spacetimes, which are save against the occurrence CTCs, won't be applicable for QFTs on spacetimes 
containing CTCs. Therefore, it is not even clear what structural properties a QFT on a spacetime containing CTCs should have, and 
given the lack of a generally accepted guideline to that effect, one could take quite diverse positions. A position one could take 
may aim to make the local algebras of observables a priori as large as possible and to view consistency conditions arising from the 
presence of CTCs as a means to classify various possiblities for CTCs to arise. A contrasting position may consist in trying to 
generalize the principles of QFT on globally hyperbolic spacetimes in such a way to quantum field theory on more general spacetimes
that the consistency conditions arising from the presence of CTCs render any quantum field theory on such spacetimes inconsitent (inexistent).
The latter view has essentially been adopted in the publications \cite{Hawking,KayFlocality,KayRadWald} to which we refer for considerable further discussion.
See also \cite{Lobo-FTP,Krasnikov-FTP,EarmanSmeenkWuethrich} for complementary views. 
However, the task we have set ourselves --- to investigate the D-CTC condition within 
a framework that refers to the principles of special or general relativity --- is without the mentioned difficulties as long as we
stay within the established approach of quantum field theories on globally hyperbolic spacetimes.

\section{The D-CTC condition in QFT on globally hyperbolic spacetimes}

Now we assume that we are given a globally hyperbolic
spacetime $M$ with metric $g$ together with an
arbitrily localizable QFT given by a collection of operator algebras $\mathcal{A}(O)$
complying with Isotony and Causality where the $O$ range over all open, relatively compact subsets of $M$.
\\[6pt]
Bipartite systems are represented by pairs operator algebras $\mathcal{A}(O_A)$ and $\mathcal{A}(O_B)$ for 
causally separated spacetime regions $O_A$ and $O_B$ as illustrated in Fig.\ 3 (see \cite{VerchWerner,HollandsSanders}).
\begin{center}
 \includegraphics[width=9cm]{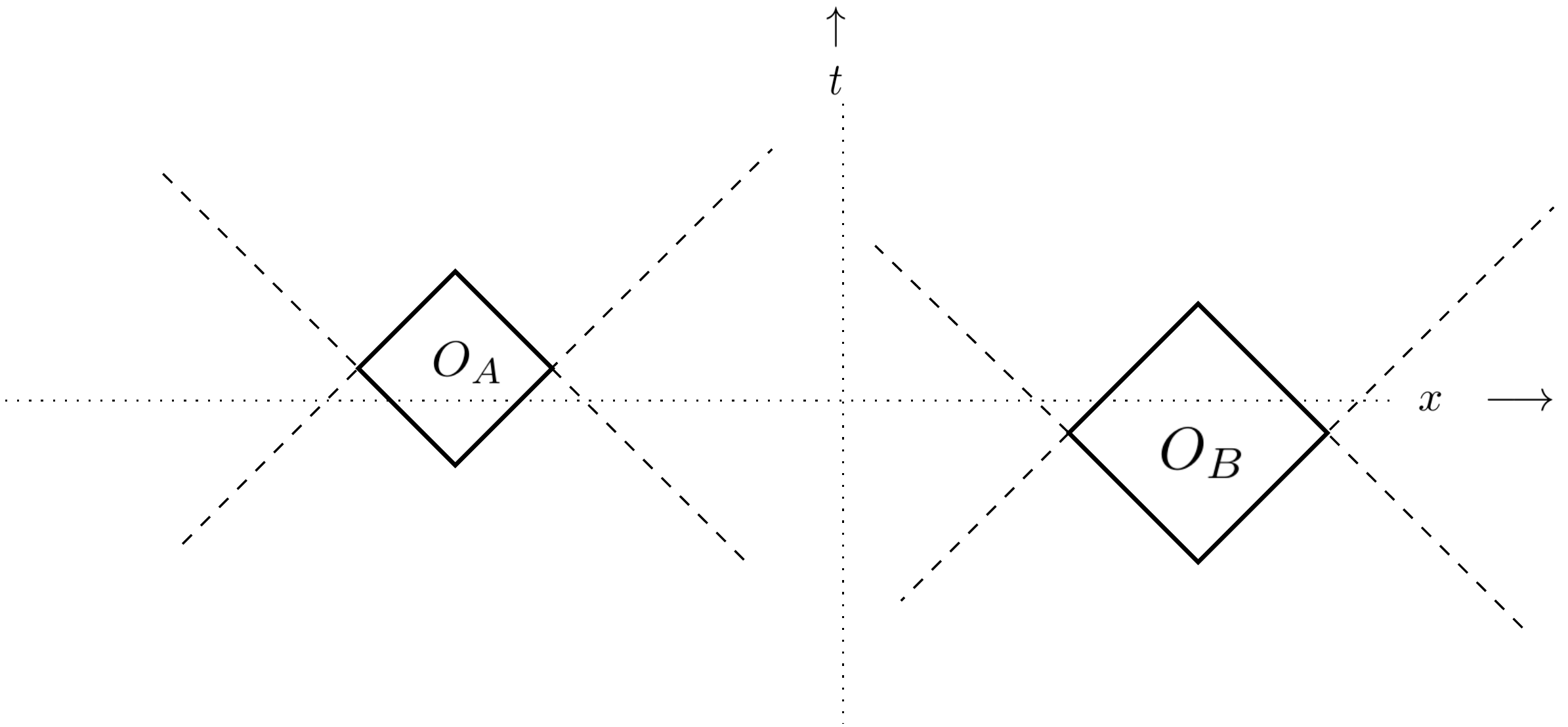}
\end{center}
{\small {\bf Figure 3.} Sketch of two spacetime regions (in 2-dimensional spacetime) which are causally separated, i.e.\
there is no causal curve connecting the closures of the spacetime regions $O_A$ and $O_B$. These spacetime regions represent
the causal completions of the times and locations where two causally separated observers carry out measurements on their 
respective parts of the system.}
\\[6pt]
Furthermore, we add the (not needed but) technically convenient assumption that
the $\mathcal{A}(O)$ are von Neumann algebras on a Hilbert space $\mathcal{H}$. This amounts to
the choice of a Hilbert space representation of the QFT. In QFT on Minkowski spacetime, this would 
typically be the vacuum representation and a similar choice could be adopted if the spacetime is 
stationary. For a QFT on a general curved spacetime, the Hilbert space representations induced 
(by means of the GNS representation) by states fulfilling the {\it microlocal spectrum condition} are regarded 
as describing physically relevant system configurations \cite{FewVer-NecHad,Fewster-Verch-Review,HollandsWald}.
With these assumptions in place, we will now investigate the extent to which the D-CTC condition can 
be fulfilled. We put this in the following form:
\\[6pt]
{\bf D-CTC Problem}: {\it Given a unitary $U$ in $\mathcal{H}$ and a density matrix state
$$ \omega_A({\bf a}) = {\rm Tr}(\varrho_A {\bf a}) \quad \ \ ({\bf a} \in \mathcal{A}(O_A)) $$
${}$ \ \ on $\mathcal{A}(O_A)$,
\\[2pt]
is there a density matrix state $\omega({\bf c}) = {\rm Tr}(\varrho {\bf c})$ on ${\sf B}(\mathcal{H})$
whose partial state on $\mathcal{A}(O_A)$ agrees with $\omega_A$ and
which is $U$-invariant in restriction to $\mathcal{A}(O_B)$, i.e.
\begin{align*}
 \omega({\bf a}) & = \omega_A({\bf a}) \quad \ \ ({\bf a} \in \mathcal{A}(O_A)) \\[2pt]
 \omega(U^\ast {\bf b} U) & = \omega({\bf b}) \quad  \ \ ({\bf b} \in \mathcal{A}(O_B)) \ \ \text{?}
\end{align*}}
Our first result displays conditions under which the D-CTC problem has no positive answer. To this end, we 
make the assumption that the spacetime is {\it ultrastatic}, i.e.\ it takes the form
$M = \mathbb{R} \times \Sigma$ with $g = dt^2 - h^{(\Sigma)}$ where $\Sigma$ is a $d$-dimensional manifold
with a complete Riemannian metric $h^{(\Sigma)}$. Thus, there is a global time coordinate $t$ 
ranging over a ``time axis'' $\mathbb{R}$, and the spacetime metric has a very rigid time-symmetry with respect to this 
time-coordinate. We furthermore suppose that
\begin{itemize}
 \item the QFT admits a time-shift symmetry implemented by  a continuous unitary group $V_t = {\rm e}^{itH}$
 with generating Hamilton operator $H \ge 0$, such that $V_t \mathcal{A}(O)V_t^* = \mathcal{A}(O_t)$, where
 $O_t = \{ (t' + t,\sigma') : (t',\sigma') \in O \subset \mathbb{R} \times \Sigma \}$
 \item 
 the QFT admits a vacuum vector, i.e.\ a unit vector $\Omega \in \mathcal{H}$ so that $V_t \Omega = \Omega$ 
 \item the QFT fulfills additivity: if a family $\{O_\nu\}$ of open, relatively compact subsets of $M$ covers $O$, then
 the von Neumann algebra generated by all the $\mathcal{A}(O_\nu)$ contains $\mathcal{A}(O)$.
 \item the QFT fulfills the timelike tube property:
 the von Neumann algebra generated by all the 
 $V_t\mathcal{A}(O)V_t^*$, $t \in \mathbb{R}$, coincides with ${\sf B}(\mathcal{H})$.
\end{itemize}
The timelike tube theorem can be seen as a variant of the Reeh-Schlieder property (see below) which holds under very general conditions
given the other assumptions \cite{Borchers-ttt,Strohmaier}. The above conditions are fulfilled e.g.\ for any QFT on Minkowski spacetime 
that derives from a Wightman-type quantum field.
\\[6pt]
A density matrix state $\omega({\bf c}) = {\rm Tr}(\varrho {\bf c})$ has {\it finite energy} if 
$$ \varrho = \frac{1}{{\rm Tr}(E \tilde{\varrho} E)} E \tilde{\varrho} E\,,$$
where $E$ denotes a spectral projector of $H$ corresponding to any finite spectral interval.
A density matrix state $\omega$ is an {\it analytic state for the energy} if it is a limit of finite energy states
preserving the analyticity properties of finite energy states with respect to $V_t$. Under the stated assumptions,\footnote{For the way the Reeh-Schlieder property is formulated here, it must be assumed that 
the closure of the spacetime region $O$ has a non-void open causal complement.} any analytic state has the  
$$ \text{\it Reeh-Schlieder property:}  \quad \ \  \omega({\bf c}^*{\bf c}) = 0 \Rightarrow {\bf c} = 0 \quad ({\bf c} \in \mathcal{A}(O))\,. $$
For more on the Reeh-Schlieder property, see \cite{StreaterWightman,Haag,S-V-W,Sanders-ReehSchl}.
\\[6pt]
We can now state our first result.
\begin{theorem} 
 Assume that the spacetime is ultrastatic and that the QFT admits a ground state and fulfills additivity and the 
 timelike tube property as described above.
 \\[6pt]
 Then for given unitary $U$ in $\mathcal{H}$ and density matrix state $\omega_A$ on $\mathcal{A}(O_A)$, there can be 
 {\it no} density matrix state $\omega$ with both of the following properties:
 \begin{itemize}
  \item[(i)] ${}$ \quad  $\omega$ is a solution to the D-CTC problem 
  \item[(ii)] ${}$ \quad  Both $\omega$ and $\omega(U^\ast \, .\, U)$ are analytic for the energy
 \end{itemize}
\end{theorem}
For the proof of this statement, see \cite{jtrv}.
 Analytic states show long-range entanglement as a consequence of the Reeh-Schlieder property \cite{CliftonHalvorson,VerchWerner}. If entanglement is taken as a distinguishing feature of a 
 quantum theory (as opposed to a classical physical theory, without a priori uncertainty relations), 
 then the result might be taken as indirectly supporting the view that in a quantum theory,
 the D-CTC condition is a characteristic feature for the occurence of CTCs (since in a QFT on a spacetime without
 CTCs, entanglement and D-CTC condition are not coexistent). However, in the light of the following results,
 this position appears to us as a stretched interpretation of the counterfactual.
\\[6pt]
The next result we are going to present shows that under very general conditions, the D-CTC condition can always be fulfilled approximately,
to any required precision, in arbitrarily localizable QFTs on any globally hyperbolic spacetime. Thus, we now assume that the manifold  $M$ 
with Lorentzian metric $g$ is a generic 
globally hyperbolic spacetime --- no symmetry assumptions are being imposed. We consider an arbitrarily localizable QFT on this spacetime, again
described by a family of von Neumann algebras $\mathcal{A}(O)$ on a Hilbert space $\mathcal{H}$, with $O$ ranging over the open, relatively compact 
subsets of $M$. Isotony and locality are imposed as before. A further assumption that we add is the {\it split property}.
\\[6pt]
The QFT fulfills the {\it split property} if,
\begin{itemize}
 \item ${}$ \ \ given a pair of stricly causally separated spacetime regions $O_A$ and $O_B$, and
 \item ${}$ \ \ given a pair of density matrix states 
 $$ \omega_A({\bf a}) = {\rm Tr}(\varrho_A{\bf a}) \quad ({\bf a} \in \mathcal{A}(O_A)) \,, \quad 
 \omega_B({\bf b}) = {\rm Tr}(\varrho_B{\bf b}) \quad ({\bf b} \in \mathcal{A}(O_B))\,,$$
\end{itemize}
there is a density matrix state $\omega({\bf c}) = {\rm Tr}(\varrho{\bf c})$ on ${\sf B}(\mathcal{H})$
which extends the two partial states as a {\it product state} (uncorrelated state), i.e.
$$ \omega({\bf ab}) = \omega_A({\bf a})\cdot \omega_B({\bf b}) \quad ({\bf a} \in \mathcal{A}(O_A)\,,\ {\bf b} \in \mathcal{A}(O_B)) \,. $$
We have used here the term ``strictly causally separated'' in order to emphasize that the {\it closures} of the spacetime regions 
$O_A$ and $O_B$ need to be causally separated which implies that the  closures of the regions cannot touch (however this agrees with our previous definition of causally separated spacetime regions
in this paper). The split property means that the subsystems of the QFT experimentally accessible within the spacetime regions $O_A$ and
$O_B$ can be made statistically independent, or isolated from one another, by preparing suitable physical states once the spacetime regions
are strictly causally separated.
The split property has been proved to hold for free QFTs and generalized free fields with a sufficiently regular mass spectrum;
it has also been proved for interacting QFTs in 1 + 1 dim. It is expected to hold generally for physically interpretable QFTs.
For considerable further discussion, see the publications \cite{BuWich,FewsterSplit,Summers-Indep,HollandsSanders}
and references cited therein.
\\[6pt]
We can now state our second result.
\begin{theorem}
 Assume that the QFT is on a general globally hyperbolic spacetime and fulfills the split property.
 Then, given any unitary $U$ in $\mathcal{H}$ and any density matrix state $\omega_A({\bf a}) = {\rm Tr}(\varrho_A{\bf a})$ on 
 $\mathcal{A}(O_A)$, there is an approximate solution to the D-CTC problem in the following sense:
 \\[6pt]
 Given arbitrary $\varepsilon > 0$ (small) and $R > 0$ (large), there is a density matrix state $\omega = \omega_{R,\varepsilon}$ on
 ${\sf B}(\mathcal{H})$ such that 
 \begin{itemize}
  \item $\omega({\bf a}) = \omega_A({\bf a})$ \quad $({\bf a} \in \mathcal{A}(O_A))$
  \item $| \omega(U^\ast {\bf b} U) - \omega({\bf b})| < \varepsilon \quad ({\bf b} \in \mathcal{A}(O_B)\,, \ ||{\bf b}|| < R)$
 \end{itemize}
\end{theorem}
For the proof, see again \cite{jtrv}. It follows from the proof that the state $\omega$ which provides an approximate solution to the 
D-CTC problem is highly non-unique; moreover, it is classically correlated (i.e.\ not entangled) across the bipartite system defined 
by the local algebras of observables $\mathcal{A}(O_A)$ and $\mathcal{A}(O_B)$. This is considerably different from our first result,
where states with a high degree of entanglement have been shown not to provide {\it exact} solutions to the D-CTC problem. 
However, if an approximate solution to the D-CTC problem is allowed to be approximate on the $\mathcal{A}(O_A)$ part of the QFT as well,
then there are also states which are analytic in the energy as solutions to the D-CTC problem. This is the statement of our next observation,
which appears here for the first time.
\\[6pt]
We again invoke the assumptions made for Theorem 1, that is, the spacetime $M$ with Lorentzian metric $g$ is assumed to be ultrastatic,
and the QFT is assumed to possess an associated time symmetry implemented by a continuous unitary group $V_t$ $(t \in \mathbb{R})$ on the 
Hilbert space $\mathcal{H}$ with
a generating Hamilton operator $H$ whose spectrum is non-negative. Furthermore, it will be assumed that there is a vacuum vector $\Omega$,
that the system of local observable algebras of the QFT fulfills additivity, and that the timelike tube property holds. In addition, we 
assume that the split property holds.
\begin{theorem}
 Let $U$ be a unitary operator on $\mathcal{H}$ and $\omega_A$ a density matrix state. Furthermore, let $\varepsilon > 0$ and $R > 0$ be 
 given. Then there is a density matrix state $\omega^{\rm an}$ on ${\sf B}(\mathcal{H})$ which is analytic in the energy and has the property 
 that 
 $$ \left| \omega_A({\bf a}) - \omega^{\rm an}({\bf a}) \right| + \left| \omega^{\rm an}(U^*{\bf b}U) - \omega^{\rm an}({\bf b}) \right| < \varepsilon $$
 holds for all ${\bf a} \in \mathcal{A}(O_A)$ and ${\bf b} \in \mathcal{A}(O_B)$ which fulfill the bound $|| {\bf a}|| + ||{\bf b}|| < R$.
\end{theorem}
{\it Proof}. Let $E_n$ be the spectral projectors of the Hamilton operator $H$ corresponding to the spectral intervals $[0,n]$
$(n \in \mathbb{N})$. Furthermore, if $\varrho$ is any density matrix on $\mathcal{H}$, define the sequence of density matrices 
\begin{align} \label{star}
 \varrho_n =\frac{1}{{\rm Tr}(E_n \varrho E_n)} E_n \varrho E_n\,.
\end{align}
Clearly, each $\varrho_n$ induces a density matrix state which is analytic in the energy.
It is plain to check that, given $\varepsilon' > 0$ and $R' > 0$, there is $n' \in \mathbb{N}$ so that 
\begin{align} \label{star2}
 \max_{||{\bf c}|| < R'} \, \left | {\rm Tr}(\varrho {\bf c}) - {\rm Tr}(\varrho_n {\bf c})\right| < \varepsilon' \quad ({\bf c} \in {\sf B}(\mathcal{H}))
\end{align}
whenever $n \ge n'$.
\\[6pt]
Now let $U$, $\omega_A$, $\varepsilon$ and $R$ be given as in the statement of the theorem. Then use Theorem 2 to conclude that there is a 
density matrix state $\omega$ with the properties 
\begin{align} \label{star3}
 \omega({\bf a}) = \omega_A({\bf a})\,, \quad | \omega(U^\ast {\bf b} U) - \omega({\bf b})| < \varepsilon /4
\end{align}
for all ${\bf a} \in \mathcal{A}(O_A)$ and all ${\bf b} \in \mathcal{A}(O_B)$ with $||{\bf b}|| < R$.
Now $\omega({\bf c}) = {\rm Tr}(\varrho {\bf c})$ $({\bf c} \in {\sf B}(\mathcal{H}))$ with a suitable density matrix $\varrho$ on
$\mathcal{H}$. Then let $\varrho_n$ be defined as in \eqref{star}. As observed above, one may choose $n'$ large enough so that
\eqref{star2} holds with $\varepsilon' = \varepsilon/4$ and $R' = R$. Then setting $\omega^{\rm an}({\bf c}) = {\rm Tr}(\varrho_{n'} {\bf c})$
$({\bf c} \in {\sf B}(\mathcal{H}))$, $\omega^{\rm an}$ is a density matrix state which is analytic in the energy. Moreover, if 
$ {\bf a} \in \mathcal{A}(O_A)$ and ${\bf b} \in \mathcal{A}(O_B)$ fulfill $||{\bf a}|| + ||{\bf b}|| < R$, combining \eqref{star2} and \eqref{star3}
yields
\begin{align*}
 | \omega_A({\bf a}) - \omega^{\rm an}({\bf a})|  \le | \omega_A({\bf a}) - \omega({\bf a})| + |\omega({\bf a}) - \omega^{\rm an}({\bf a})| 
   < \varepsilon/4
\end{align*}
and
\begin{align*}
 |\omega^{\rm an}(U^*{\bf b}U) - \omega^{\rm an}({\bf b})| & \le |\omega(U^*{\bf b}U) - \omega({\bf b})| + 
     |\omega(U^*{\bf b}U) - \omega^{\rm an}(U^*{\bf b}U)| \\ & \ + |\omega({\bf b}) - \omega^{\rm an}({\bf b})| \\
     & \ < \varepsilon/4 + \varepsilon/4 + \varepsilon/4 = \frac{3}{4}\varepsilon\,.
\end{align*}
This proves the statement of the theorem. \hfill $\Box$

\section{Discussion}

We have related the D-CTC condition to spacetime structure using the local, operator algebraic framework of relativistic 
quantum field theory. In doing so, Theorem 1 puts constraints on the exact fulfillment of the D-CTC condition in the 
presence of entanglement. However, Theorems 2 and 3 show that the D-CTC condition can always be approximately fulfilled 
to any precision
in quantum field theory on globally hyperbolic spacetimes where there are no closed timelike curves. These results should 
be taken as indicating that any interpretation of the D-CTC condition beyond a mere analogy to closed timelike curves in
the sense of general relativity is potentially misleading and ought to be regarded with caution. The D-CTC condition makes no reference to locality and 
causality concepts and is therefore not sufficiently specific to be in any sense characteristic for quantum processes in
the presence of closed timelike curves in the sense of general relativity. As an interesting note, there is a contrast in 
the results of Theorem 1 and Theorem 3 depending on whether the D-CTC condition is required to be fulfilled exactly or 
approximately. This indicates that the level of mathematical formalization and idealization of the D-CTC condition, or 
other potential conditions delineating the behaviour of processes in the presence of closed timelike curves in quantum field theory,
needs to be considered very carefully. We hope that this will be discussed profoundly in future investigations.

\end{document}